\documentclass[aps,twocolumn,superscriptaddress,eqsecnum,nofootinbib,showpacs,
preprintnumbers ] {
revtex4}
\usepackage{graphicx,epsfig}
\usepackage{amsmath}
\usepackage {amssymb}

\newcommand{\be}{\begin{eqnarray}}
\newcommand{\ee}{\end{eqnarray}}
\newcommand{\bea}{\begin{eqnarray}}
\newcommand{\nn}{\nonumber}
\newcommand{\eea}{\end{eqnarray}}

\def\a{\alpha}
\def\b{\beta}

\def\d{\delta}

\def\la{\lambda}

\def\k{\kappa}
\def\m{\mu}
\def\n{\nu}

\def\f{\phi}

\def\z{\zeta}

\begin{document}

\title{Cosmology in new gravitational scalar-tensor theories}

\author{Emmanuel N. Saridakis}
\email{Emmanuel_Saridakis@baylor.edu}
\affiliation{CASPER, Physics Department, Baylor University, Waco, TX 76798-7310, USA}
\affiliation{Instituto de F\'{\i}sica, Pontificia Universidad de Cat\'olica de 
Valpara\'{\i}so, 
Casilla 4950, Valpara\'{\i}so, Chile}

\author{Minas Tsoukalas}
\email{minasts@central.ntua.gr}
 \affiliation{Physics Department, Bo\u{g}azi\c{c}i University, 
34342, Bebek, Istanbul, Turkey}


\begin{abstract}
 We investigate the cosmological applications of new gravitational scalar-tensor 
theories, which are novel modifications of gravity possessing $2+2$ propagating 
degrees of freedom, arising from a Lagrangian that includes the Ricci scalar and its first 
and second derivatives. Extracting the field equations we obtain an effective dark energy 
sector that consists of both extra scalar degrees of freedom, and we determine 
various observables. We analyze two specific models and we obtain a cosmological behavior 
in agreement with observations, i.e. transition from matter to dark energy 
era, with the onset of cosmic acceleration. Additionally, for a particular 
range of the model parameters, the equation-of-state parameter of the effective dark 
energy sector can exhibit the phantom-divide crossing. These features reveal the 
capabilities of these theories, since they arise solely from the novel, 
higher-derivative terms.
\end{abstract}


\maketitle

\section{Introduction}

According to the standard model of cosmology, which is supported by a huge amount of 
observations, the expansion of the universe includes two accelerated phases, at early and 
late times respectively. Since this behavior cannot be described within the standard  
paradigm of physics, namely within general relativity and Standard Model of particles, 
physicists try to increase the degrees of freedom of the theory. In principle there are 
two ways to achieve it. The first is to modify the universe content by introducing new, 
exotic, fields, such as the inflaton \cite{Olive:1989nu,Bartolo:2004if} or the concept of 
dark energy \cite{Copeland:2006wr,Cai:2009zp}. The second is to consider that the extra 
degrees of freedom are gravitationally oriented, i.e. that they arise from a 
gravitational modification at specific scales \cite{Nojiri:2006ri,Capozziello:2011et}. 
Note that the second approach, apart from the above cosmological motivation, has a 
theoretical motivation too, namely to improve the UltraViolet behavior of gravity 
\cite{Stelle:1976gc,Biswas:2011ar}. Finally, we mention that the above constructions 
are not separated by strict boundaries, since one can completely or partially 
transform between them, or build theories where both extensions are used.

In order to construct a gravitational modification one can add higher-order corrections 
to the action of general relativity, as in $f(R)$ gravity 
\cite{Sotiriou:2008rp,DeFelice:2010aj,Nojiri:2010wj,Nojiri:2006gh}, 
in Gauss-Bonnet and $f(G)$ gravity \cite{Nojiri:2005jg,DeFelice:2008wz}, in Lovelock 
gravity \cite{Lovelock:1971yv,Deruelle:1989fj}, in Weyl gravity
\cite{Mannheim:1988dj,Flanagan:2006ra}, etc. However, one should ensure himself that the 
additional degrees of freedom introduced in such modifications do not present a ghost 
behavior or other kind of catastrophic instabilities, at the background or perturbation 
levels. Indeed, Horndeski was able to construct the most general single-scalar field 
theory with second-order equations of motion and thus without ghosts in 
\cite{Horndeski:1974wa}, a construction which was re-discovered in the framework of  
Galileon modifications in 
\cite{Nicolis:2008in,Deffayet:2009wt,Deffayet:2009mn,Deffayet:2011gz}. These classes of 
theories involve $2+1$ propagating degrees of freedom, that is one extra comparing to 
general relativity (see also the extension to beyond Horndeski theories, with still one 
extra propagating degree of freedom 
\cite{Zumalacarregui:2013pma,Gleyzes:2014dya,Gao:2014soa,Gleyzes:2014qga,Gao:2014fra}).

Hence, a question arises naturally: can we construct gravitational modifications beyond 
the above classes, i.e. possessing for instance  $2+2$ propagating degrees of freedom, 
while still being ghost-free? Indeed, such a multi-scalar modification was developed in
\cite{Padilla:2010de,Padilla:2012dx}, and it was shown to be the most general 
multi-scalar tensor theory in a flat background \cite{Sivanesan:2013tba} but not in 
a general one \cite{Kobayashi:2013ina}. Nevertheless, the construction of the most 
general field equations for a bi-scalar \cite{Ohashi:2015fma} or multi-scalar theory still 
attracts a lot of interest in the literature \cite{Padilla:2013jza}, as well as the 
corresponding cosmological and black-hole applications 
\cite{Padilla:2010tj,Charmousis:2014zaa}. 

However, although the construction of ghost-free theories with $2+2$ or more propagating 
degrees of freedom is obtained relatively easily in the scalar-field language, it proves 
to be a harder task if one starts from the pure gravitational modification formulation. 
Indeed, in \cite{Naruko:2015zze} the authors managed to construct a modified gravity 
using the Ricci scalar and its first and second derivatives, which under a specific 
Lagrangian choice is free of ghosts, possessing $2+2$ degrees of freedom, namely 2 scalar 
degrees and 2 tensor ones. These constructions, named gravitational scalar-tensor 
theories, are equivalent with specific cases of generalized bi-Galileon theories, however 
whether there is a complete one-to-one correspondence between them is an open question.

In the present work we are interested in investigating the cosmological behavior in  
gravitational scalar-tensor theories. In particular, we desire to study the late-time 
evolution of a universe governed by such a gravitational modification, and examine 
whether we can obtain acceleration without the use of an explicit cosmological constant, 
namely arising solely from the novel  terms of the theory. The plan of the work is as 
follows: In Section \ref{themodel} we present the new gravitational scalar-tensor 
theories and we derive the corresponding field equations in a general background. Then, 
in Section \ref{Cosmology} we apply them in a cosmological framework, and we explicitly 
investigate two specific models. Finally, Section \ref{Conclusions} is devoted to 
summary and discussion.

\section{New gravitational scalar-tensor theories}
\label{themodel}

In this section we briefly review the construction of gravitational scalar-tensor 
theories following \cite{Naruko:2015zze}, starting from the modified gravitational 
action and resulting in the corresponding specific bi-scalar action. Then we derive the 
general equations of motion for both metric and scalar degrees of freedom, in a general 
background.
 
The starting point for the construction of new gravitational scalar-tensor theories is 
the idea to (re-)formulate generalized scalar-tensor theories only in terms of the metric 
and its derivatives, without the use of a scalar field \cite{Naruko:2015zze}.  Hence, one 
starts by extending the $f(R)$ action to include derivatives of the Ricci scalar, namely
\be
\label{bfR}
S=\int d^{4}\sqrt{-g}\, f\left(R,(\nabla R)^{2},\square R \right),
\ee
where $(\nabla R)^{2}=g^{\m\n}\nabla_{\m}R\nabla_{\n}R$. These actions, despite their 
higher derivative nature, using double Lagrange multipliers can be transformed to actions 
of multi-scalar fields coupled minimally to gravity. A crucial step is the dependence of 
$f$ on $\square R=\b$. In the case where it does not enter linearly, namely if 
$f_{\b\b}\neq0$, where subscripts denote partial derivatives, then (\ref{bfR}) can be 
re-written in the following form:
\begin{eqnarray}
&&\!\!\!\!\!\!\!\!
S=\!\int \!
d^{4}x\sqrt{-\hat{g}}\left\{\frac{1}{2}\hat{R}-\frac{1}{2}\hat{g}^{\m\n}\left(\partial_{\m
}
\chi\partial_{\n}\chi+e^{-\sqrt{\frac{2}{3}}\chi}\partial_{\m}\f\partial_{\n}\f 
\right)\right.\nonumber\\
&& \left.
-\frac{1}{4}\!\left[e^{-\sqrt{\frac{2}{3}}\chi}\f+e^{-2\sqrt{\frac{2}{3}}\chi}
\left(\varphi\,\b(\f,(\hat{ \nabla}\f)
^{2},\varphi)\!-\!f\right) \right]\! \right\}\!,
\end{eqnarray}
where $\chi,\f$ are scalar fields and  $\varphi\equiv f_{\b}$ (for simplicity, here and 
in the following, we set the gravitational constant to one). In the above expression the
 $hat$ denotes a frame conformally related to the original one through 
$g_{\mu\nu}=\frac{1}{2}e^{-\sqrt{\frac{2}{3}}\chi}\hat{g}_{\mu\nu}$.

If on the other hand 
$\b$ does enter linearly, namely if $f_{\b\b}=0$, then the function $f$ can be re-written 
as
\begin{equation}
f(R,(\nabla R)^{2},\square R)=\mathcal{K}((R,(\nabla R)^{2})+\mathcal{G}(R,(\nabla 
R)^{2})\square R.
\end{equation}
If $\mathcal{G}$ depends only on $R$, then through integration by parts the second term of 
the above relation can be redefined in terms of the first one, and 
thus this case is equivalent to the $\mathcal{G}=0$ one. However, in the general case 
where $\mathcal{G}=\mathcal{G}(R,(\nabla R)^{2})$ the above new gravitational 
scalar-tensor theory can be transformed to the following bi-scalar construction 
\cite{Naruko:2015zze}:
\begin{eqnarray}
\label{action}
&&\!\!\!\!\!\!\!\!\!\!\!\!\!\!\!
S=\int d^{4}x 
\sqrt{-\hat{g}}\left[\frac{1}{2}\hat{R}-\frac{1}{2}\hat{g}^{\m\n}\nabla_{\m}\chi\nabla_{\n
}\chi\right.\nonumber\\
&& \ \ \ \ \ \ \ \ \ \ \ \
- 
\frac{1}{\sqrt{6}}e^{-\sqrt{\frac{2}{3}}\chi}\hat{g}^{\m\n}\mathcal{G}\nabla_{\m}
\chi\nabla_{\n}\phi+\frac{1}{4}e^{
-2\sqrt{
\frac{2}{3}}
\chi}\mathcal{K}
\nonumber\\
&& \ \ \ \ \ \ \ \ \ \ \ \
\left.
+\frac{1}{2}e^{-\sqrt{\frac{2}{3}}\chi}\mathcal{G}\hat{\square}\phi-\frac{1}{4}e^{-\sqrt{
\frac{2}{3}}\chi}\phi\right],
\end{eqnarray}
where now
\be
\mathcal{K}=\mathcal{K}(\phi,B),\quad\mathcal{G}=\mathcal{G}(\phi,B),
\ee
 with
\be
B=2e^{\sqrt{
\frac{2}{3}}\chi}g^{\m\n}\nabla_{\m}\phi\nabla_{\n}\phi.
\ee

The above action contains two scalar fields, namely $\chi$ and $\f$, however it does so 
in the specific and suitable combination in order to be equivalent with the original 
higher-derivative gravitational action. Hence, although in simple $f(R)$ theories the 
conformal transformation leads to the replacement of the functional degree of freedom of 
$f(R)$ by a scalar field, in the above constructions the derivatives of $R$ are not 
replaced by derivatives of the scalar-field in a naive way, but only through the above 
two-field combination. That is why the authors of \cite{Naruko:2015zze} named these 
theories as ``new gravitational scalar-tensor theories'', in a sense that they can be 
understood as the pure gravitational formulations of standard multi-scalar-tensor 
theories constructed from scalar fields and the metric. Such theories have not been 
previously investigated and thus they may open new paths towards the construction of 
gravitational modifications.

In the following we will work with the Einstein-frame version of the above 
theories, namely with action (\ref{action}), and thus for simplicity we drop the hats.
Varying (\ref{action}) with respect to the metric leads to the metric field equations 
\begin{eqnarray}
\label{metricfieldeq}
\mathcal{E}_{\m\n}&\!\!=\!\!&\frac{1}{2}G_{\m\n}+\frac{1}{4}g_{\m\n}g^{\a\b}\nabla_{\a}
\chi\nabla_
{\b}\chi-\frac{1}{2}\nabla_{\m}\chi\nabla_{\n}\chi\nonumber\\
&&
+\frac{1}{4}g_{\m\n}\sqrt{\frac{2}{3}}e^
{-\sqrt{ \frac{2}{3}}
\chi}g^{\a\b}\mathcal{G}\nabla_{\a}\chi\nabla_{\b}\phi\nonumber\\
&&
-\frac{1}{2}\sqrt{\frac{2}{3}}e^{
-\sqrt{\frac{
2}{3}}\chi}\mathcal{G}\nabla_{(\m}\chi\nabla_{\n)}\phi \nn\\
&&-\sqrt{\frac{2}{3}}g^{\a\b}\nabla_{\a}\chi\nabla_{\b}\phi\,\mathcal{G}_{B}\nabla_{\m}
\phi\nabla_{\n}\phi
\nonumber\\
&&
-\frac{1}{4}g_{\m\n}e^{-\sqrt{\frac{2}{3}}\chi}\mathcal{G}
\square\phi+\mathcal{G}_{B 
}(\square\phi)\nabla_{\m}\phi\nabla_{\n}\phi
\nonumber\\
&&
+\frac{1}{2}e^{-\sqrt{\frac{2}{3}}\chi}
\mathcal{G} \nabla_{\m}\nabla_{\n}\phi 
-\frac{1}{2}\nabla_{\k}\left(e^{-\sqrt{\frac{2}{3}}\chi}\mathcal{G}\d^{\la}_{(\m}\d^{\k}
_{\n)} 
\nabla_{\la}\phi\right)
\nonumber\\
&&
+\frac{1}{4}\nabla_{\k}\left(e^{-\sqrt{\frac{2}{3}}\chi}\mathcal{G}
g_{\m\n}\nabla^{\k}\phi\right)-\frac{1}{8}g_{\m\n}e^{-2\sqrt{\frac{2}{3}}\chi}\mathcal{K}
\nonumber\\
&&
+\frac{1}{ 2}e^{-\sqrt{\frac{2}{3}}\chi}\mathcal{K}_{B}\nabla_{\m}\phi\nabla_{\n}\phi 
+\frac{1}{8}g_{\m\n}e^{-\sqrt{\frac{2}{3}}\chi}\phi=0,
\end{eqnarray}
where the parentheses in space-time indices mark symmetrization, and the subscripts in 
$\mathcal{G}$ and $\mathcal{K}$  denote partial derivatives with respect to the 
corresponding 
argument (for instance  $\mathcal{G}_{B}=\frac{\partial \mathcal{G}(\phi,B)}{\partial B}$
etc). Similarly, variation of (\ref{action}) with respect to the scalar fields $\chi$ 
and $\phi$ leads respectively to their equations of motion, namely
\begin{eqnarray}
\mathcal{E}_{\chi}&\!\!=\!\!&\square\chi+\frac{1}{3}e^{-\sqrt{\frac{2}{3}}\chi}g^{\m\n}
\mathcal{ G}
\nabla_{\m}\chi\nabla_{\n}\phi\nonumber\\
&&
-\frac{2}{3}g^{\m\n}\nabla_{\m}\chi\nabla_{\n}\phi\,\mathcal
{G}_{B}g^ 
{\a\b}\nabla_{\a}\phi\nabla_{\b}\phi\nonumber\\
&&
+\frac{1}{2}\sqrt{\frac{2}{3}}\nabla_{\m}\left( 
e^{-\sqrt{\frac{2}{3}}\chi}g^{\m\n}\mathcal{G}\nabla_{\n}\phi\right)\nn\\
&&-\frac{1}{2}\sqrt{\frac{2}{3}}e^{-\sqrt{\frac{2}{3}}\chi}\mathcal{G}\square\phi+\sqrt{
\frac{2}{3}}
\mathcal{G}_{B}\nabla_{\m}\phi\nabla_{\n}\phi\,g^{\m\n}\square\phi
\nonumber\\
&&
-\frac{1}{2}\sqrt{\frac{
2}{3}}e^{-
2\sqrt{\frac{2}{3}}\chi}\mathcal{K}+\frac{1}{2}e^{-\sqrt{\frac{2}{3}}\chi}\mathcal{K}_{B}
\sqrt{\frac{2}{3}}g^{\m\n}\nabla_{\m}\phi\nabla_{\n}\phi
\nn\\
&&+\frac{1}{4}\sqrt{\frac{2}{3}}e^{-\sqrt{\frac{2}{3}}\chi}\phi=0,
\label{chifieldeq}
\end{eqnarray}
and 
\begin{eqnarray}
\mathcal{E}_{\phi}&\!\!=\!\!&-\frac{1}{2}\sqrt{\frac{2}{3}}e^{-\sqrt{\frac{2}{3}}\chi}g^{
\m\n }
\mathcal{G}_{\phi}\nabla_{\m}\chi\nabla_{\n}\phi
\nonumber\\
&&
+2\sqrt{\frac{2}{3}}\nabla_{\b}\left(g^{
\m\n}\mathcal{G} 
_{B}g^{\a\b}\nabla_{\a}\phi\nabla_{\m}\chi\nabla_{\n}\phi\right)
\nonumber\\
&&
+\frac{1}{2}\sqrt{\frac{2}
{3}} 
\nabla_{\n}\left(e^{-\sqrt{\frac{2}{3}}\chi}g^{\m\n}\mathcal{G}\nabla_{\m}\chi\right)\nn\\
&&+\frac{1}{2}e^{-\sqrt{\frac{2}{3}}\chi}\mathcal{G}_{\phi}\square\phi-2\mathcal{G}_{B}
(\square\phi)
^{2}-2\nabla_{\n}\mathcal{G}_{B}\square\phi\nabla^{\n}\phi
\nonumber\\
&&
-\frac{1}{2}\sqrt{\frac{2}{3}}
\nabla^{\m}\left(e^{-\sqrt{\frac{2}{3}}\chi}\nabla_{\m}\chi\,\mathcal{G} 
\right)+\frac{1}{2}\nabla^{\m}\left(e^{
-\sqrt{\frac{2}{3}}\chi}\,\mathcal{G}_{\phi}\nabla_{\m}\phi\right)\nn\\
&&-\frac{1}{2}\sqrt{\frac{2}{3}}e^{-\sqrt{\frac{2}{3}}\chi}\nabla^{\m}\chi\mathcal{G}_{B}
\nabla_{\m}
B+\frac{1}{2}e^{-\sqrt{\frac{2}{3}}\chi}\nabla^{\m}\mathcal{G}_{B}\nabla_{\m}B
\nonumber\\
&&
+\sqrt{\frac{2}{3}}
e^{-\sqrt{
\frac{2}{3}}\chi}\mathcal{G}_{B}\nabla^{\m}\left(e^{\sqrt{\frac{2}{3}}
\chi}\nabla_{\m}\chi\nabla^{\n}\phi\nabla_{\n}\phi\right)\nn\\
&&
+2 e^{-\sqrt{\frac{2}{3}}\chi}\mathcal{G}_{B}  
\nabla^{\m}\left(e^{\sqrt{\frac{2}{3}}\chi}\nabla^{
\n}\phi\right)\nabla_{\m}\nabla_{\n}\phi
\nonumber\\
&&
+2\mathcal{G}_{B}R^{\m\n}\nabla_{\m}\phi\nabla_{\n
}\phi+\frac{1}{4}e^{-2\sqrt{\frac{2}{3}}\chi}\mathcal{K}_{\phi}\nn\\
&&-\nabla_{\n}\left(e^{-\sqrt{\frac{2}{3}}\chi}\mathcal{K}_{B}g^{\m\n}\nabla_{\m}
\phi\right)-\frac{
1}{4}e^{-\sqrt{\frac{2}{3}}\chi}=0.
\label{phifieldeq}
\end{eqnarray}
We stress here that despite the fact that the higher-order term in the action 
(\ref{action}) naively seems to be problematic, in the sense that it could lead to 
higher-order derivatives in the field equation for $\phi$, it proves to be perfectly 
fine, just like the corresponding term in simple Horndeski theory 
\cite{Horndeski:1974wa,Deffayet:2011gz}. Lastly, note that the scenario at hand reproduces 
standard general relativity in the case where $\mathcal{K}=\phi/2$ and $\mathcal{G}=0$, 
and the triviality of the conformal transformation in this case leads to 
$\chi=-\sqrt{\frac{3}{2}}\ln2$.

\section{Cosmology in new gravitational scalar-tensor theories}
\label{Cosmology}

The new gravitational scalar-tensor theories presented above are novel gravitational 
modifications, and hence it would be interesting to examine their cosmological 
applications. The first thing one should do in order to investigate the cosmology in a 
universe governed by such gravitational theories is to introduce the matter content. 
Although incorporating the matter sector in the original Jordan frame or in the Einstein 
one would lead to different theories, in this work we prefer for simplicity to introduce 
it straightaway in the action (\ref{action}), leaving the alternative approach for a 
future study. Hence, we consider the total action $S_{tot}=S+S_{m}$, and thus the metric  
field equations (\ref{metricfieldeq}) now become
\be
\mathcal{E}_{\m\n}=\frac{1}{2}T_{\m\n},
\ee
where $T_{\m\n}=\frac{-2}{\sqrt{-g}}\frac{\d S_{m}}{\d g^{\m\n}}$ is the energy-momentum 
tensor of the matter perfect fluid. Furthermore, we consider a flat 
Friedmann-Robertson-Walker (FRW) spacetime metric of the form
 \be
ds^{2}=-dt^{2}+a(t)^{2} \delta_{ij}dx^{i}dx^{j},
\ee
where $a(t)$ is the scale factor, and thus the two scalars are time-dependent only. Under 
these considerations, the metric field equations (\ref{metricfieldeq}) give rise to the 
Friedmann equations, namely
\begin{eqnarray}
&&\!\!\!\!\!\!\!\!\!\!\!\!\!\!\!\!\!
3H^{2}-\rho_m-\frac{1}{2}\dot{\chi}^{2}+\frac{1}{4}e^{-2\sqrt{\frac{2}{
3}}\chi}\mathcal{K}
\nonumber\\
&& \!\!\!\!
+\frac{2}{3}\dot{\phi}^{2}\left[\dot{\phi}\left(\sqrt{6}\dot{\chi}
-9H\right)-3\ddot{\phi} 
\right]\mathcal{G}_{B}
\nn\\
&&  \!\!\!\!
-\frac{1}{2}e^{-\sqrt{\frac{2}{3}}\chi}\left[\dot{B}\dot{\f}\mathcal{
G}_{B}+ \frac{\f}{2} +\dot{\f}^{2}\left(\mathcal{G}_{\f}-2\mathcal{K}_{B} \right) 
 \right]\! =0,  
\label{FR1}
\end{eqnarray}
\begin{eqnarray}
&&\!\!\!\!\!\!\!\!\!\!\!\!\!\!\!\!\!\!\!\!\!\!
3H^{2}+2\dot{H}+p_m+\frac{1}{2}\dot{\chi}^{2}+\frac{1}{4}e^{-2\sqrt{
\frac{2}{3}}
\chi}\mathcal{K}
\nonumber\\
&&\!\!\!\!\!\!\!\!\!
+\frac{1}{2} e^{-\sqrt{\frac{2}{3}}\chi}\left(-\frac{\f}{2}+\dot{B}\dot{\f
}\mathcal{
G}_{B}+\dot{\f}^{2}\mathcal{G}_{\f} \right)=0,
  \label{FR2}
\end{eqnarray}
where now $B(t)=2 e^{\sqrt{\frac{2}{3}}\chi} g^{\m\n}\nabla_{\m}\f\nabla_{\n}\f=-2 
e^{\sqrt{\frac{2}{3}}\chi} \dot{\f}^{2}$,  $H=\dot{a}/a$ is the Hubble 
parameter, and a dot denotes differentiation with respect to $t$. In these expressions  
$\rho_m$ and $p_m$ are respectively the energy density and pressure of the matter fluid.
Similarly, from the two scalar field equations (\ref{chifieldeq}) and (\ref{phifieldeq}) 
we respectively obtain their evolution equations, namely
\begin{eqnarray}
&&\!\!\!\!\!\!\!\!\!\!\!\!\!\!
\mathcal{E}_{\chi}=\ddot{\chi}+3H\dot{\chi}-\frac{1}{3}\dot{\f}^{2}\left[\dot{\f}
\left(3\sqrt{6}H-
2\dot{\chi} \right)+\sqrt{6}\ddot{\f} \right]\mathcal{G}_{B}\nn\\
&&+
\frac{1}{
2\sqrt{6}}
e^{-\sqrt{\frac{2}{3}}\chi}\left[2\dot{B}\dot{\f}\mathcal{G}_{B}-\f+2\dot{\f}^{2}
\left(\mathcal{K}_{
B}+\mathcal{G}_{\f} \right) \right]\nonumber\\
&&+\frac{1}{\sqrt{6}}e^{-2\sqrt{\frac{2}{3}}\chi}\mathcal{K}
=0,
\label{chiequation}
\end{eqnarray}
and
\begin{eqnarray}
&&\!\!\!\!\!\!\!\!\! \!
\mathcal{E}_{\phi}=
\frac{1}{3}e^{-\sqrt{\frac{2}{3}}\chi}
\left[\dot{\f}\left(-9H+\sqrt{6}\dot{\chi}
\right)-3\ddot{\f}\right]\mathcal{K}_{B}
\nn
\\
&&
+\frac{1}{6}\dot{B}\left\{3e^{-\sqrt{\frac{2}{3}}
\chi}\dot{B} +4\dot{\f}
\left[\dot{\f}\left(9H-\sqrt{6}\dot{\chi}\right)+3\ddot{\f}\right]\right\}\mathcal{G}_{BB}
\nn\\
&&
+\frac{1}{3}e^{-\sqrt{
\frac{2}{3}
}\chi}\left[\dot{\f}\left(9H-\sqrt{6}\dot{\chi}\right)+3\ddot{\f}\right]\mathcal{G}_{\f}
\nn
\\
&&
+\left\{e^{-\sqrt{\frac{2}{3}}\chi}\dot{B}\dot{\f}+\frac{2}{3}\dot{\f}^{2}
\left[\dot{\f}\left(9H-\sqrt{6}\dot{\chi}\right)+3\ddot{\f}\right]\right\}\mathcal{G}_{B 
\f}
\nn
\\
&&
-e^{-\sqrt{\frac{2}{3}}\chi}\dot{\f}^{2}\mathcal{K}_{B 
\f}+\frac{1}{2}e^{-\sqrt{\frac{2}{3}}\chi}\dot{\f}^{2}\mathcal{G}_{\f\f}
 -e^{-\sqrt{\frac{2}{3}}\chi}\dot{B}\dot{\f}\mathcal{K}_{BB}
 \nn
\\
&&
+\left[
 \frac{4}{3}\dot{\f}\left(9H-2\sqrt{6}\dot
{\chi} 
\right)\ddot{\f}
-\frac{1}{\sqrt{6}}e^{
-\sqrt{\frac{2}{3}}\chi}\dot{B}\dot{\chi}
\right.\nn\\
&&\left. \ \ \ \,
+\dot{\f}^2\left(18H^{2}+6\dot{H}-3\sqrt{6}H\dot{\chi}-\frac{2}{3}\dot{\chi}^{2}
-\sqrt{6} \ddot{\chi}\right)\right]
\mathcal{G}_{B}
\nn
\\
&&
-\frac{1}{4}
e^{-2\sqrt{\frac{2}{3}}\chi}\mathcal{K}_{\f}+ \frac{1}{4}e^{-\sqrt{\frac{2}{3}}\chi}
=0,
\label{phiequation}
\end{eqnarray}
where we use the notation $\mathcal{G}_{B \f}=\mathcal{G}_{\f 
B}=\frac{\partial^{2}\mathcal{G}}{\partial B \partial \f}$, etc. One can rigorously  
verify that the above equations are compatible with the equations 
arising from the fact that the total action is diffeomorphism  invariant 
\cite{Ohashi:2015fma}:
\be
\nabla_\m \mathcal{E}^{\m\n}+\frac{1}{2}\mathcal{E}_{\chi} 
\nabla^{\n}\chi+\frac{1}{2}\mathcal{E}_{\f} 
\nabla^{\n}\phi=\frac{1}{2}\nabla_{\m}T^{\m\n}=0.
\ee 
 
Concerning the late-time application of the above equations, one can see that the 
Friedmann equations 
(\ref{FR1}),(\ref{FR2}) can be written in the usual form, namely
\begin{eqnarray}
&&H^2=\frac{1}{3}(\rho_{DE}+\rho_m)
 \label{FR1b}
 \\
&&2\dot{H}+3H^2=-(p_{DE}+p_m),
 \label{FR2b}
\end{eqnarray}
 defining
an effective dark energy sector with energy density and pressure respectively as:
\begin{eqnarray}
  \label{rhoDE}
 &&\!\!\!\!\!\!\!\!\!\!\!\!\!\!\!\!\!\!\!\!
 \rho_{DE}\equiv 
 \frac{1}{2}\dot{\chi}^{2}-\frac{1}{4}e^{-2\sqrt{\frac{2}{
3}}\chi}\mathcal{K}
\nn\\
&&\!\!
-\frac{2}{3}\dot{\phi}^{2}\left[\dot{\phi}\left(\sqrt{6}\dot{\chi}
-9H\right)-3\ddot{\phi} 
\right]\mathcal{G}_{B}
\nn\\
&&\!\!
+\frac{1}{2}e^{-\sqrt{\frac{2}{3}}\chi}\left[\dot{B}\dot{\f}\mathcal{
G}_{B}+ \frac{\f}{2} +\dot{\f}^{2}\left(\mathcal{G}_{\f}-2\mathcal{K}_{B} \right)
 \right],
 \end{eqnarray}
\begin{eqnarray}
\label{pDE}
&&\!\!\!\!\!\!\!\!\!\!\!\!\!\!\!\!\!\!\!\!\!\!\!\!\!\!\!\!\!\!\!\!\!\!\!\!\!\!\!\!\!
p_{DE}\equiv  
 \frac{1}{2}\dot{\chi}^{2}+\frac{1}{4}e^{-2\sqrt{
\frac{2}{3}}
\chi}\mathcal{K}
\nn\\
&&\!\!\!\!\!\!\!\!\!\!\!\!\!\!\!\!\!\!\!\!\!\!\!
+
\frac{1}{2} e^{-\sqrt{\frac{2}{3}}\chi}\left(\dot{B}\dot{\f
}\mathcal{
G}_{B}+\dot{\f}^{2}\mathcal{G}_{\f}-\frac{\f}{2} \right).
\end{eqnarray}
Therefore, in the new gravitational scalar-tensor theories at hand, we obtain an 
effective dark-energy sector that consists of both extra scalar degrees of freedom. 
Using the scalar field equations of motion (\ref{chiequation}) and (\ref{phiequation}), 
one can straightforwardly see that
\begin{equation}
\dot{\rho}_{DE}+3H(\rho_{DE}+p_{DE})=0,
\end{equation}
while the corresponding dark-energy equation-of-state parameter is given by: 
\begin{equation}
\label{wDE}
w_{DE}\equiv \frac{p_{DE}}{\rho_{DE}}.
\end{equation}
Finally, note that the matter energy density and pressure satisfy the standard
evolution equation 
\begin{equation}
\label{rhoevol}
\dot{\rho}_m+3H(\rho_m+p_m)=0.
\end{equation}

In order to examine the cosmological application of the above construction, we have to 
consider specific ansatzes for the functions $\mathcal{K}(\phi,B)$ and 
$\mathcal{G}(\phi,B)$. In the following subsections we consider two of 
such examples, corresponding to the first non-trivial extensions of general relativity 
possessing $2+2$ degrees of freedom.

\subsection{Model 1: $\mathcal{K}(\f,B)=\frac{\f}{2}-\frac{\z}{2}B$ and 
$\mathcal{G}(\f,B)=0$}

As a first example let us investigate the case where 
\be
\mathcal{K}(\f,B)=\frac{1}{2}\f-\frac{\z}{2}B \quad \text{and} \quad \mathcal{G}(\f,B)=0,
\ee
with $\zeta$ the corresponding coupling constant. We remind that in FRW geometry we have  
  $B(t)=-2 
e^{\sqrt{\frac{2}{3}}\chi} \dot{\f}^{2}$.
In this case the Friedmann equations  (\ref{FR1}),(\ref{FR2}) read
\begin{equation}
\label{Fr1rv}
3 
H^{2}-\rho_m-\frac{1}{2}\dot{\chi}^{2}+\frac{1}{8}e^{-2\sqrt{\frac{2}{3}}\chi}\f-\frac{1}{
4 }
e^{-\sqrt{\frac{2}{3}}\chi}\left(\f+\z\dot{\f}^{2} \right)=0,
\end{equation}
\bea
&&\!\!\!\!\!\!\!\!\!\!\!\!\!\!\!\!\!\!\!\!\!\!\!\!\!\!\!\!\!\!\!\!\!\!\!\!\!\!\!\!\!\!\!\!
\!
3 H^{2}+2 
\dot{H}+p_m+\frac{1}{2}\dot{\chi}^{2}+\frac{1}{8}e^{-2\sqrt{\frac{2}{3}}\chi}\f
\nonumber\\
&&\!\!
-\frac{1}{4}
e^{-\sqrt{\frac{2}{3}}\chi}\left(\f-\z\dot{\f}^{2} \right)=0,
\eea
while   the two scalar field equations (\ref{chiequation}) and (\ref{phiequation}) write 
as
\begin{equation}
\ddot{\chi}+3 H 
\dot{\chi}+\frac{1}{2\sqrt{6}}e^{-2\sqrt{\frac{2}{3}}\chi}\f-\frac{1}{2\sqrt{6}}e^{-
\sqrt{\frac{2}{3}}\chi}\left(\f-\z\dot{\f}^{2} \right)=0,
\end{equation}
\begin{equation}
\z \ddot{\phi}+\frac{1}{3}\z\dot{\f}\left(9 
H-\sqrt{6}\dot{\chi}\right)-\frac{1}{4}e^{-\sqrt{\frac{
2}{3}}\chi}+\frac{1}{2}=0.
\end{equation}
Hence, in this case the effective dark-energy energy density and pressure 
(\ref{rhoDE}),(\ref{pDE}) become
\begin{equation}
  \label{rhoDE2}
 \rho_{DE}= 
 \frac{1}{2}\dot{\chi}^{2}-\frac{1}{8}e^{-2\sqrt{\frac{2}{3}}\chi}\f+\frac{1}{4}
e^{-\sqrt{\frac{2}{3}}\chi}\left(\f+\z\dot{\f}^{2} \right),
 \end{equation}
\begin{equation}
\label{pDE2}
p_{DE}=
 \frac{1}{2}\dot{\chi}^{2}+\frac{1}{8}e^{-2\sqrt{\frac{2}{3}}\chi}\f-\frac{1}{4}
e^{-\sqrt{\frac{2}{3}}\chi}\left(\f-\z\dot{\f}^{2} \right).
\end{equation}

The above equations do not accept analytical solutions, and hence in order to 
investigate the cosmological evolution we perform a numerical elaboration. We consider 
the matter sector to be dust, and thus we set $p_m\approx0$. Additionally, in order 
to acquire a consistent cosmology in agreement with observations, we set the present 
values of the density parameters to $\Omega_{m0}=\frac{\rho_{m0}}{3H^2}\approx0.3$ and
$\Omega_{DE0}=\frac{\rho_{DE0}}{3H^2}\approx0.7$ \cite{Ade:2013sjv}. Finally, as usual it 
proves convenient to use the  redshift $z=-1+a_0/a$ as the independent variable, setting 
the current scale factor  $a_0$ to 1.

\begin{figure}[ht]
\begin{center}
\includegraphics[width=0.5\textwidth]{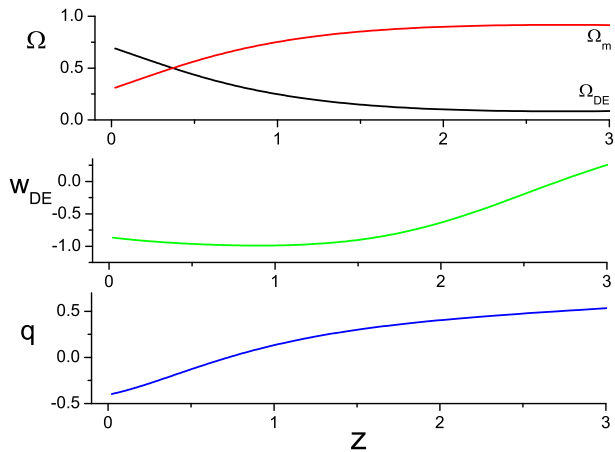}
\caption{{\it{The late-time cosmological evolution, for Model 1, for
the parameter choice $\zeta=10$ (in units where the gravitational constant is set to 
one), having imposed $\Omega_{m0}\approx0.3$,
$\Omega_{DE0}\approx0.7$ at present, and having set the present
scale factor $a_0=1$. As independent variable we use the redshift
$z=-1+a_0/a$. In the upper graph we depict the evolution of the
matter and dark energy density parameters. In the middle graph we
present the evolution of the dark-energy equation-of-state
parameter $w_{DE}$. In the lower graph we depict the evolution of
the deceleration parameter $q$.}}} \label{Model1DE}
\end{center}
\end{figure}

In Fig. \ref{Model1DE} we present the cosmological evolution for the parameter choice 
$\zeta=10$, focusing on various observables. In particular, in the upper graph we depict 
the evolution of the matter and dark energy density parameters, and as we observe it is 
in agreement with the observed one \cite{Ade:2013sjv}. In the middle graph of Fig. 
\ref{Model1DE} we depict the evolution of the dark-energy equation-of-state
parameter $w_{DE}$. As we can see, it presents a dynamical behavior, and at late times it 
almost stabilizes in a value very close to the cosmological-constant one, as expected 
from observations. Finally, in the  lower graph of Fig. \ref{Model1DE} we depict the 
evolution of the deceleration parameter $q=-1-\dot{H}/H^2$, where we can clearly see the 
passage from deceleration ($q>0$) to acceleration ($q<0$) in the recent cosmological 
past, as it is required from observations. 

In summary, the cosmological behavior of the scenario at hand is in agreement with 
observations. We stress here that we have not considered an explicit cosmological 
constant, and thus the obtained acceleration arises solely from the novel, 
higher-derivative terms of the gravitational scalar-tensor theories.

\begin{figure}[ht]
\begin{center}
\includegraphics[width=0.5\textwidth]{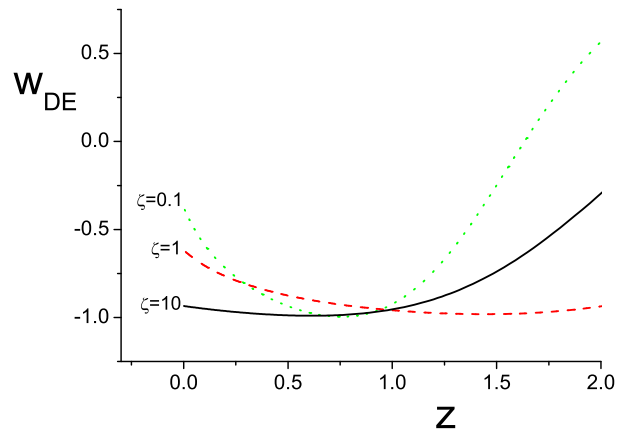}
\caption{{\it{The evolution of the dark-energy equation-of-state parameter $w_{DE}$ as a 
function of the redshift $z=-1+a_0/a$, for six values of the model parameter $\zeta$ (in 
units where the gravitational constant is set to one), having imposed 
$\Omega_{m0}\approx0.3$, $\Omega_{DE0}\approx0.7$ at present, and having set the present
scale factor $a_0=1$.  }}} \label{Model1w}
\end{center}
\end{figure}

Let us now investigate how the model parameter $\zeta$ affects the behavior of the 
dark-energy equation-of-state parameter $w_{DE}$. In Fig. \ref{Model1w} we present the 
evolution of $w_{DE}$ for various values of $\zeta$ (we consider $\zeta>0$ in order for 
the $\phi$ field not to exhibit an effective ghost behavior in (\ref{Fr1rv})). As we 
observe, $w_{DE}$ lies in the quintessence regime and with 
increasing $\zeta$ its final value comes closer to the cosmological-constant value $-1$.

\subsection{Model 2: $\mathcal{K}(\f,B)=\frac{\f}{2}$ and 
$\mathcal{G}(\f,B)=\xi B$}

As a second example we consider the case where
\be
\mathcal{K}(\f,B)=\frac{1}{2}\f \quad \text{and} \quad \mathcal{G}(\f,B)=\xi B,
\ee
with $\xi$ the corresponding coupling constant (in FRW geometry we have  $B(t)=-2 
e^{\sqrt{\frac{2}{3}}\chi} \dot{\f}^{2}$).
Thus, the Friedmann equations  (\ref{FR1}),(\ref{FR2}) become
\bea
\label{Fr1mod2rv}
&&\!\!\!\!\!\!\!\!\!\!\!\!\!\!\!\!\!
3 H^{2}-\rho_m-\frac{1}{2}\dot{\chi}^{2}+\frac{1}{8}e^{-2\sqrt{\frac{2}{3}}\chi}\left(1-2 
e^{\sqrt{\frac{2}{3}}\chi} 
\right)\f\nonumber\\
&& \ \ \ \ \ \,
+\xi\dot{\f}^{3}\left(\sqrt{6}\dot{\chi}-6H 
\right)=0,
\eea
\bea
&&\!\!\!\!\!\!\!\!\!\!\!\!\!\!\!\!\!
3 H^{2}+2 
\dot{H}+p_m+\frac{1}{2}\dot{\chi}^{2}+\frac{1}{8}e^{-2\sqrt{\frac{2}{3}}\chi}\left(1-2 
e^{\sqrt{\frac{2}{3}}\chi} 
\right)\f
\nonumber\\
&& \ \ \  \ \ \,
-\frac{1}{3}\xi\dot{\f}^{2}\left(\sqrt{6}\dot{\f}\dot{\chi}+6\ddot{\f} 
\right)=0,
\eea
while the two scalar field equations (\ref{chiequation}) and (\ref{phiequation}) read as
\bea
&&\!\!\!\!\!\!\!\!\!\!\!\!\!\!\!\!\!\!\!\!\!\!\!\!\!
\ddot{\chi}+3 H 
\dot{\chi}+\frac{1}{2\sqrt{6}}e^{-2\sqrt{\frac{2}{3}}\chi}\left(1-e^{\sqrt{\frac{2}{
3}}\chi}\right)\f\nonumber\\
&&
-\sqrt{6}\,\xi\dot{\f}^{2}\left(H\dot{\f}+\ddot{\f} \right)=0,
\eea
\bea
&&\!\!\!\!\!\!\!\!\!\! 
\xi\dot{\f}\left\{2\left(-6H+\sqrt{6}\dot{\chi}\right)\ddot{\f}\right.\nonumber\\
&&\left.
+\dot{\f}\left[-6\dot{H}
+3H\left(-6H+\sqrt{6}\dot{\chi}\right)+\sqrt{6}\ddot{\chi}\right]\right\}\nonumber\\
&&\!\!\!\!\!\!\!\!\!
+\frac{1}{8}e^{
-2\sqrt{\frac{2}{ 3}}\chi}\left(1-2 e^{\sqrt{\frac{2}{3}}\chi} \right)=0.
\eea
Therefore, in this case the effective dark-energy energy density and pressure 
(\ref{rhoDE}),(\ref{pDE}) write as
\begin{eqnarray}
  \label{rhoDE3}
 &&\!\!\!\!\!\!\!\!\!\!\!\!\!\!\!\!\!\!
 \rho_{DE}= 
 \frac{1}{2}\dot{\chi}^{2}-\frac{1}{8}e^{-2\sqrt{\frac{2}{3}}\chi}\left(1-2 
e^{\sqrt{\frac{2}{3}}\chi} \right)\f
\nonumber\\
&&
-\xi\dot{\f}^{3}\left(\sqrt{6}\dot{\chi}-6H 
\right),
 \end{eqnarray}
\begin{eqnarray}
\label{pDE3}
 &&\!\!\!\!\!\!\!\!\!\!\!\!\!\!\!\!\!\!
p_{DE}=
\frac{1}{2}\dot{\chi}^{2}+\frac{1}{8}e^{-2\sqrt{\frac{2}{3}}\chi}\left(1-2 
e^{\sqrt{\frac{2}{3}}\chi} 
\right)\f
\nonumber\\
&&
-\frac{1}{3}\xi\dot{\f}^{2}\left(\sqrt{6}\dot{\f}\dot{\chi}+6\ddot{\f} 
\right).
\end{eqnarray}

\begin{figure}[!]
\begin{center}
\includegraphics[width=0.5\textwidth]{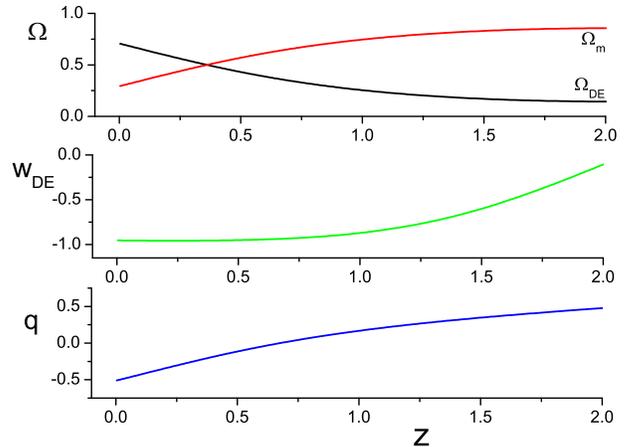}
\caption{{\it{The late-time cosmological evolution, for Model 2, for
the parameter choice $\xi=-1$ (in units where the gravitational constant is set to 
one), having imposed $\Omega_{m0}\approx0.3$,
$\Omega_{DE0}\approx0.7$ at present, and having set the present
scale factor $a_0=1$. As independent variable we use the redshift
$z=-1+a_0/a$. In the upper graph we depict the evolution of the
matter and dark energy density parameters. In the middle graph we
present the evolution of the dark-energy equation-of-state
parameter $w_{DE}$. In the lower graph we depict the evolution of
the deceleration parameter $q$.}}} \label{Model2DE}
\end{center}
\end{figure}

In Fig. \ref{Model2DE} we depict the cosmological evolution for the parameter choice 
$\xi=-1$. In the upper graph we show the behavior of the matter and dark energy density 
parameters, which is in agreement with the observed one \cite{Ade:2013sjv}. In the middle 
graph of Fig. \ref{Model2DE} we depict the evolution of the dark-energy equation-of-state
parameter $w_{DE}$, which presents a dynamical behavior, and at late times it 
almost stabilizes in a value very close to the cosmological-constant one, as expected 
from observations. Finally, in the lower graph of Fig. \ref{Model2DE} we present the 
evolution of the deceleration parameter, from which we can see the 
passage from deceleration  to acceleration. Similarly to Model 1 of the previous 
subsection, we mention that the onset of acceleration is a pure effect of the novel, 
higher-derivative terms of the gravitational scalar-tensor theories.

\begin{figure}[!]
\begin{center}
\includegraphics[width=0.5\textwidth]{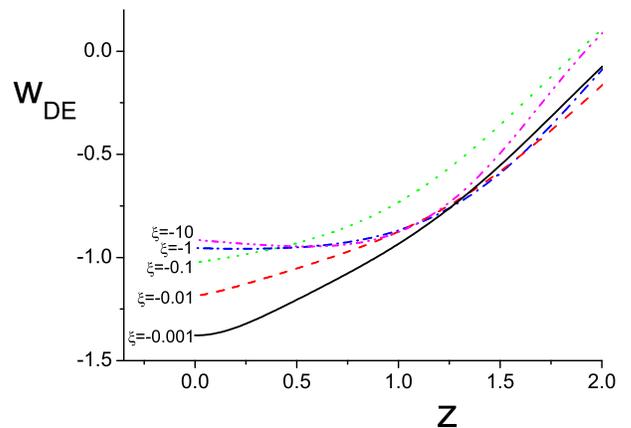}
\caption{{\it{The evolution of the dark-energy equation-of-state parameter $w_{DE}$ as a 
function of the redshift $z=-1+a_0/a$, for five values of the model parameter $\xi$ (in 
units where the gravitational constant is set to one), having imposed 
$\Omega_{m0}\approx0.3$,$\Omega_{DE0}\approx0.7$ at present, and having set the present
scale factor $a_0=1$.  }}} 
\label{Model2w}
\end{center}
\end{figure}

In order to see how the model parameter $\xi$ affects the behavior of $w_{DE}$, in Fig. 
\ref{Model2w} we present the evolution of $w_{DE}$ for various values of $\xi$. As we 
can see, for large negative $\xi$-values $w_{DE}$ lies in the quintessence regime, 
while for small negative values it exhibits the phantom-divide crossing and lies below 
$-1$ at current times. Although this phantom behavior might be a signal that the $\phi$ 
field behaves effectively as a ghost, this does not need necessarily to be the case since 
$\phi$-field's effective kinetic energy in (\ref{Fr1mod2rv}) has a complicated form 
depending on the time-derivatives of both fields as well as of the scale factor, and thus 
the phantom behavior can result even if the fields behave as canonical ones 
\cite{Nojiri:2013ru}. Clearly, the safe procedure to investigate this issue is to 
perform a detailed hamiltonian analysis, a task that lies beyond the scope of the present 
work and thus it is left for a future project.

\section{Conclusions}
\label{Conclusions}

New gravitational scalar-tensor theories are novel modifications of gravity possessing  
$2+2$ propagating degrees of freedom. Although similar models had been 
constructed in the scalar-field language, for instance in bi-scalar or bi-Galileon models, 
it is not straightforward to develop them in the pure gravitational language. However, 
such a construction is indeed possible using the Ricci scalar and its first and second 
derivatives under a specific Lagrangian that is free of ghosts \cite{Naruko:2015zze}. The 
crucial point is that although in simple $f(R)$ theories the conformal transformation 
leads to the replacement of the functional degree of freedom of $f(R)$ by a scalar field, 
in the above constructions the derivatives of $R$ are not replaced by derivatives of the 
scalar-field in a naive way, but only through a specific and suitable two-field 
combination. 

In this work we investigated the cosmological applications of new gravitational 
scalar-tensor theories. Introducing the matter sector and considering a homogeneous and 
isotropic geometry, we extracted the Friedmann equations, as well as the evolution 
equations of the new extra scalar degrees of freedom. In such a scenario, we obtain an 
effective dark energy sector that consists of both extra scalar degrees of freedom, and 
hence we can determine various observables, such as the dark-energy and matter density 
parameters, the dark-energy equation-of-state parameter and the deceleration parameter.
 
We analyzed two specific models, corresponding to the first non-trivial extensions of 
general relativity possessing $2+2$ degrees of freedom. As we showed, the resulting 
cosmological behavior is in agreement with observations, i.e. we obtain the transition 
from the matter to the dark energy era, with the onset of cosmic acceleration. Moreover,  
the equation-of-state parameter of the effective dark energy sector can be stabilized in a 
value very close to the cosmological-constant one. The most interesting feature is that 
such a behavior arises solely from the novel, higher-derivative terms of the gravitational 
scalar-tensor theories, since we have not considered an explicit cosmological constant. 
Additionally, we saw that for a particular range of the model parameters, the dark-energy
equation-of-state parameter can exhibit the phantom-divide crossing in the recent 
cosmological past and currently lie in the phantom regime, which might be the case 
according to observations. This feature reveals the capabilities of new gravitational 
scalar-tensor theories, since the phantom behavior could be obtained even if the fields 
behave as canonical ones.

The above features indicate that the new gravitational scalar-tensor theories provide an 
interesting candidate for modified theories of gravity. Hence it would be worthy to 
perform detailed 
investigations on their applications. Firstly, one should perform a detail confrontation 
with observational data from Type Ia Supernovae (SNIa), Baryon Acoustic Oscillations 
(BAO), and Cosmic Microwave Background (CMB) observations, to constrain the possible 
classes of such modifications. Additionally, one should perform a complete phase-space 
analysis, in order to extract information about the global late-time behavior of the above 
scenarios. Moreover, an extensive analysis of the perturbations is a necessary task that 
could bring these constructions closer to detailed data such as those related to the 
growth index and the large-scale structure. Furthermore, one should examine the black hole 
solutions in the framework of new gravitational scalar-tensor theories, in order to 
obtain additional information on their novel features. Finally, one could try to analyze 
further extensions along this direction, using for instance terms of the form 
$(\nabla_{\m}\nabla_{\n}R)^{2}=\nabla_{\m}\nabla_{\n}R\nabla^{\m}\nabla^{\n}R$. These 
projects are left for near-future investigations.

 \begin{acknowledgments} 
MT is supported by TUBITAK 2216 fellowship under the application number 1059B161500790.
\end{acknowledgments}



\begin{thebibliography}{99}


 
\bibitem{Olive:1989nu} 
  K.~A.~Olive,
  Phys.\ Rept.\  {\bf 190}, 307 (1990).

 
\bibitem{Bartolo:2004if} 
  N.~Bartolo, E.~Komatsu, S.~Matarrese and A.~Riotto,
  Phys.\ Rept.\  {\bf 402}, 103 (2004).


 
\bibitem{Copeland:2006wr}
  E.~J.~Copeland, M.~Sami and S.~Tsujikawa,
  Int.\ J.\ Mod.\ Phys.\  D {\bf 15}, 1753 (2006).

 
 

\bibitem{Cai:2009zp}
  Y.~F.~Cai, E.~N.~Saridakis, M.~R.~Setare and J.~Q.~Xia,
  Phys.\ Rept.\  {\bf 493}, 1 (2010).

 

 
 
 
\bibitem{Nojiri:2006ri}
  S.~Nojiri and S.~D.~Odintsov,
  eConf {\bf C0602061}, 06 (2006), Int.\ J.\ Geom.\ Meth.\ Mod.\ Phys.\ 
{\bf 4}, 115 (2007).


\bibitem{Capozziello:2011et}
  S.~Capozziello and M.~De Laurentis,
  Phys.\ Rept.\  {\bf 509}, 167 (2011).
 
 
 
 
  \bibitem {Stelle:1976gc}
  K.~S.~Stelle,
Phys.\ Rev.\ D \textbf{16}, 953 (1977).
 
\bibitem{Biswas:2011ar} 
  T.~Biswas, E.~Gerwick, T.~Koivisto and A.~Mazumdar,
  Phys.\ Rev.\ Lett.\  {\bf 108}, 031101 (2012).

\bibitem{Sotiriou:2008rp} 
  T.~P.~Sotiriou and V.~Faraoni,
  Rev.\ Mod.\ Phys.\  {\bf 82}, 451 (2010).


\bibitem{DeFelice:2010aj}
  A.~De Felice and S.~Tsujikawa,
  Living Rev.\ Rel.\  {\bf 13}, 3 (2010).
 

  
    
\bibitem{Nojiri:2010wj} 
  S.~'i.~Nojiri and S.~D.~Odintsov,
  Phys.\ Rept.\  {\bf 505}, 59 (2011).
 
 
  
  
 
 
  
\bibitem{Nojiri:2006gh} 
  S.~Nojiri and S.~D.~Odintsov,
  Phys.\ Rev.\ D {\bf 74}, 086005 (2006).
 
 
 
 


\bibitem{Nojiri:2005jg}
  S.~'i.~Nojiri and S.~D.~Odintsov,
  Phys.\ Lett.\ B {\bf 631}, 1 (2005).
 
 
 


\bibitem{DeFelice:2008wz}
  A.~De Felice and S.~Tsujikawa,
  Phys.\ Lett.\ B {\bf 675}, 1 (2009).
 
 
\bibitem{Lovelock:1971yv}
  D.~Lovelock,
  J.\ Math.\ Phys.\  {\bf 12}, 498 (1971).




\bibitem{Deruelle:1989fj}
  N.~Deruelle and L.~Farina-Busto,
  Phys.\ Rev.\ D {\bf 41}, 3696 (1990).



\bibitem{Mannheim:1988dj}
  P.~D.~Mannheim and D.~Kazanas,
  Astrophys.\ J.\  {\bf 342}, 635 (1989).

  
 \bibitem{Flanagan:2006ra}
  E.~E.~Flanagan,
  Phys.\ Rev.\ D {\bf 74}, 023002 (2006).





\bibitem{Horndeski:1974wa}
  G.~W.~Horndeski,
  Int.\ J.\ Theor.\ Phys.\  {\bf 10} (1974) 363.
  
  
\bibitem{Nicolis:2008in}
  A.~Nicolis, R.~Rattazzi and E.~Trincherini,
  Phys.\ Rev.\ D {\bf 79} (2009) 064036.
  




\bibitem{Deffayet:2009wt} 
  C.~Deffayet, G.~Esposito-Farese and A.~Vikman,
  Phys.\ Rev.\ D {\bf 79}, 084003 (2009).

  
 
 \bibitem{Deffayet:2009mn} 
  C.~Deffayet, S.~Deser and G.~Esposito-Farese,
  Phys.\ Rev.\ D {\bf 80}, 064015 (2009).





  
  
  
  
  
\bibitem{Deffayet:2011gz}
  C.~Deffayet, X.~Gao, D.~A.~Steer and G.~Zahariade,
  Phys.\ Rev.\ D {\bf 84} (2011) 064039.

  
\bibitem{Zumalacarregui:2013pma} 
  M.~Zumalacárregui and J.~García-Bellido,
  Phys.\ Rev.\ D {\bf 89}, 064046 (2014).
  
\bibitem{Gleyzes:2014dya} 
  J.~Gleyzes, D.~Langlois, F.~Piazza and F.~Vernizzi,
  Phys.\ Rev.\ Lett.\  {\bf 114}, no. 21, 211101 (2015).

\bibitem{Gao:2014soa} 
  X.~Gao,
  Phys.\ Rev.\ D {\bf 90}, 081501 (2014).


\bibitem{Gleyzes:2014qga} 
  J.~Gleyzes, D.~Langlois, F.~Piazza and F.~Vernizzi,
  JCAP {\bf 1502}, 018 (2015).

  
\bibitem{Gao:2014fra} 
  X.~Gao,
  Phys.\ Rev.\ D {\bf 90}, 104033 (2014).

  

\bibitem{Padilla:2010de}
  A.~Padilla, P.~M.~Saffin and S.~Y.~Zhou,
  JHEP {\bf 1012} (2010) 031.

\bibitem{Padilla:2012dx}
  A.~Padilla and V.~Sivanesan,
  JHEP {\bf 1304} (2013) 032.

\bibitem{Sivanesan:2013tba}
  V.~Sivanesan,
  Phys.\ Rev.\ D {\bf 90} (2014) 10,  104006.
 
 
\bibitem{Kobayashi:2013ina}
  T.~Kobayashi, N.~Tanahashi and M.~Yamaguchi,
  Phys.\ Rev.\ D {\bf 88} (2013) 8,  083504.


  
\bibitem{Ohashi:2015fma}
  S.~Ohashi, N.~Tanahashi, T.~Kobayashi and M.~Yamaguchi,
  JHEP {\bf 1507} (2015) 008.

   
  

\bibitem{Padilla:2013jza}
  A.~Padilla, D.~Stefanyszyn and M.~Tsoukalas,
  Phys.\ Rev.\ D {\bf 89} (2014) 6,  065009.

\bibitem{Padilla:2010tj} 
  A.~Padilla, P.~M.~Saffin and S.~Y.~Zhou,
  JHEP {\bf 1101}, 099 (2011).

  
  
\bibitem{Charmousis:2014zaa}
  C.~Charmousis, T.~Kolyvaris, E.~Papantonopoulos and M.~Tsoukalas,
  JHEP {\bf 1407} (2014) 085.

\bibitem{Naruko:2015zze}
  A.~Naruko, D.~Yoshida and S.~Mukohyama,
  arXiv:1512.06977 [gr-qc].

\bibitem{Ade:2013sjv}
  P.~A.~R.~Ade {\it et al.} [Planck Collaboration],
  Astron.\ Astrophys.\  {\bf 571}, A1 (2014).
  

\bibitem{Nojiri:2013ru} 
  S.~Nojiri and E.~N.~Saridakis,
  Astrophys.\ Space Sci.\  {\bf 347}, 221 (2013).
  
  


  


  
\end{thebibliography}
\end{document}